\pdfoutput=1

\newif\ifprl
\prlfalse

\newcommand*{\ATLASLATEXPATH}{./}

\ifprl 
   \documentclass[aps,prl,reprint,USenglish,texlive=2014]{revtex4-1} 
\else
   \documentclass[cernpreprint,texlive=2016,txfonts,UKenglish]{\ATLASLATEXPATH atlasdoc}
\fi

\usepackage[biblatex=false]{\ATLASLATEXPATH atlaspackage}
\usepackage{\ATLASLATEXPATH atlasphysics}

\usepackage{HmumuPaper-defs}
\usepackage{dcolumn}
\usepackage{multirow}
\usepackage{rotating}
\usepackage{amssymb}
\usepackage{tabularx}
\usepackage{tabularew}
\usepackage{adjustbox}
\usepackage{tabu}

\ifprl
    
   \usepackage{\ATLASLATEXPATH atlasmetadefs}          
\else
   \usepackage{cite}                                   
   \PreprintIdNumber{CERN-EP-2017-078}                
   \author{The ATLAS Collaboration}
\fi

\DeclareOldFontCommand{\rm}{\normalfont\rmfamily}{\mathrm}

\hypersetup{pdftitle={ATLAS document},pdfauthor={The ATLAS Collaboration}}

\AtlasTitle{Search for the Dimuon Decay of the Higgs Boson in $pp$ Collisions at $\sqrt{s}$ = 13~\TeV\ with the ATLAS Detector}

\AtlasJournalRef{Phys. Rev. Lett. 119 (2017) 051802}
\AtlasDOI{10.1103/PhysRevLett.119.051802}

\AtlasAbstract{A search for the dimuon decay of the Higgs boson was performed using data corresponding to an
integrated luminosity of 36.1~fb$^{-1}$
collected with the ATLAS detector in $pp$ collisions at $\sqrt{s}=13$~\TeV\ at the Large Hadron Collider.
No significant excess is observed above the expected background.
The observed (expected) upper limit on the cross section times branching ratio
is 3.0~(3.1) times the Standard Model prediction at the 95\% confidence level for a Higgs boson mass of 125~\GeV.
When combined with the $pp$ collision data at $\sqrt{s}=7$~\TeV\ and $\sqrt{s}=8$~\TeV,
the observed (expected) upper limit is 2.8~(2.9) times the Standard Model prediction.
}

\begin{document}

\ifprl
    \title{\AtlasTitleText}         
    \author{The ATLAS Collaboration}
    \begin{abstract}                
    \AtlasAbstractText              
    \end{abstract}                  
\fi

\maketitle

In July 2012, the ATLAS and CMS Collaborations discovered a new particle with a mass of approximately 125~\GeV~\cite{ATLAS:2012obs,CMS:2012obs,CMSLong2013} 
at the CERN Large Hadron Collider (LHC).
Subsequent measurements have indicated that this particle is 
consistent with the Standard Model (SM) Higgs boson~\cite{ATLAS_combination,CMS_combination,Chatrchyan:2012jja,atlas_spin_paper,Khachatryan:2014kca,Khachatryan:2016vau,ATLASCMSHmass}, denoted by $H$.
The $\hmm$ decay is a sensitive channel in which the Higgs coupling to second-generation fermions can be measured 
 with a clean final-state signature at the LHC.
The SM branching ratio for the Higgs boson to dimuon decay is $2.18\times 10^{-4}$~\cite{Heinemeyer:2013tqa} for $m_H = 125$~\GeV.
 Several scenarios beyond the SM~\cite{Giudice:2008uua,Vignaroli:2009vt,Dery:2013rta} predict a higher branching ratio. 
 Any deviation from the SM prediction could be a sign of new physics. 
The ATLAS experiment carried out a search for the  $\hmm$ process 
using data collected in 2011 and 2012 (LHC Run 1), corresponding 
to integrated luminosities of 4.5~$\text{fb}^{-1}$ at a center-of-mass energy $\sqrt{s}=7$~\TeV\ and 20~$\text{fb}^{-1}$ 
at $\sqrt{s}=8$~\TeV~\cite{Aad:2014xva}. 
For a Higgs boson with a mass of 125~\GeV, an observed (expected) upper limit of 7.1~(7.2) was set at the 95\% confidence level (C.L.) on
the signal strength, defined as the production rate of the \hmm process normalized to the SM prediction.
The CMS experiment also performed searches for the $\hmm$ process with data collected in LHC Run 1~\cite{CMS:Hmumu}. 
The observed (expected) upper limit from CMS on the signal strength was 7.4~(6.5) at the 95\% C.L. for 
a Higgs boson with $m_H = 125$~\GeV.

In this Letter, a search for the dimuon decay of the Higgs boson is presented. 
The Higgs boson mass is assumed to be $m_H = 125$~\GeV\ for all the results presented in this Letter. 
The search is performed using $pp$ collision data recorded with the ATLAS detector in 2015 and 2016 at $\sqrt{s}=13$~\TeV. 
The data set corresponds to an integrated luminosity of \ulumi.
This analysis selects events with exactly two opposite-charge muons and classifies them into eight orthogonal categories. 
Two categories are defined using a multivariable discriminant and provide good sensitivity to the vector-boson fusion (VBF) process.
Signal events produced in the VBF process tend to have two high-\pT forward jets in opposite detector hemispheres
and little hadronic activity between them.  
The other six categories are sensitive to signal events produced in the gluon--gluon fusion (\textit{gg}F) process 
and are defined with different requirements on muon pseudorapidity ($\eta$) and the transverse momentum of the dimuon system (\ptmm). 
The dominant irreducible background is the $Z/\gamma^{*} \rightarrow \mu\mu$ (Drell-Yan) process. 
A simultaneous fit to distributions of the dimuon invariant mass $m_{\mu\mu}$ 
in all the categories is performed in the range 110 to 160~\GeV\ to extract the overall $\hmm$ signal strength and determine the background normalizations and shapes. 
The fitting range is chosen to avoid the $Z$ boson mass peak and have enough data events to constrain the background.

The ATLAS detector\cite{ATLASdetector} at the LHC covers nearly the entire solid angle around the collision point~\cite{AtlasCoordFootnote}.
It consists of an inner tracking detector (ID) surrounded by a thin superconducting solenoid, 
electromagnetic and hadronic calorimeters, and a muon spectrometer (MS) incorporating three large superconducting toroidal magnets.
Events used in this analysis were recorded using a combination of single-muon
triggers, with the transverse momentum (\pT) threshold being 26~\GeV\ for isolated muons 
or 50~\GeV\ for muons without any isolation requirement imposed.  
The trigger efficiency is about 95\% for the signal processes.

Monte Carlo (MC) simulated samples are used to optimize the event selection, to model the signal processes and 
to develop an analytic function to model the $m_{\mu\mu}$ distributions for the total background. 
Signal events from the \textit{gg}F and VBF processes were generated with \textsc{Powheg-box} v2~\cite{Frixione:2007vw}
at next-to-leading order (NLO) in quantum chromodynamics (QCD) using the CT10~\cite{Lai:2010vv} 
parton distribution function (PDF) set
and \textsc{Pythia8}~\cite{Sjostrand:2007gs}
for parton showering and hadronization.
\textsc{Pythia8}  was  also used to model \hmm events produced in association with a $W$ or $Z$ boson ($VH$).
The hadronization and underlying-event parameters were set according to the AZNLO tune 
based on the $Z$ boson \pT distribution measurement in 7~\TeV\ $pp$ collisions~\cite{Aad:2014xaa}. 
The simulated Higgs boson $\pt$ spectrum for the \textit{gg}F process is tuned to match the \textsc{HRes} prediction~\cite{deFlorian:2012mx,Grazzini:2013mca}.

The signal samples are normalized to the predicted cross sections times branching ratio. 
The production cross sections of the Higgs boson at $\sqrt{s}=13$~\TeV\ are reported in Refs.~\cite{Dittmaier:2011ti,Dittmaier:2012vm,Heinemeyer:2013tqa}.
The cross section for the \textit{gg}F process is calculated at next-to-next-to-next-to-leading-order QCD~\cite{Anastasiou:2016cez} 
and NLO electroweak accuracies~\cite{Aglietti:2004nj,Actis:2008ug}.
Both the VBF and $VH$ cross sections are computed with next-to-next-to-leading-order QCD~\cite{Bolzoni:2010xr} 
and NLO electroweak precision~\cite{Ciccolini:2007jr,Ciccolini:2007ec,Arnold:2008rz}.
The branching ratio for the $\hmm$ decay is calculated using \textsc{Hdecay}~\cite{Djouadi:1997yw} at NLO in QCD.

Drell-Yan background events were generated with \textsc{MadGraph}5~\cite{Alwall:2011uj} 
with the \textsc{NNPDF23LO}~\cite{Ball:2012cx} PDF set interfaced to \textsc{Pythia8}. 
The $t\bar{t}$ and single-top quark samples were generated with \textsc{Powheg-box} v2 
using the \textsc{CT10} PDF set interfaced to \textsc{Pythia6}~\cite{Sjostrand:2006za} for parton showering and hadronization.
The diboson processes ($WW$, $WZ$, and $ZZ$) were generated with \textsc{Sherpa} v2.1~\cite{Gleisberg:2008ta}  with the \textsc{CT10} PDF set.

All simulated samples were processed through the full ATLAS detector simulation\cite{Aad:2010ah} based on \textsc{Geant4}\cite{Agostinelli:2002hh}. 
The effects arising from multiple $pp$ collisions in the same or neighboring bunch crossings (pileup) were included in the MC simulation.          
%For all the samples, simulated distributions of the average number of interactions per bunch crossing are reweighted to match that in data.
Events are reweighted such that the distribution of the average number of interactions per bunch crossing matches that observed in data. 
Simulated events are corrected to reflect the muon momentum scale and resolution and the muon  trigger and identification  efficiencies measured in data.

Events are required to contain at least one reconstructed $pp$ collision vertex candidate with 
at least two associated ID tracks each with $\pT > 0.4$~\GeV. 
The vertex with the largest sum of ${\pT}^2$ of tracks is considered to be the primary vertex.
Dimuon events are selected by requiring two opposite-charge muons. 
Muons are reconstructed by combining tracks in the ID with tracks in the MS.
Candidate muons are required to satisfy the ``medium'' criteria 
defined in Ref.~\cite{Aad:2016jkr} and required to have $\pT>15$~\GeV\ and $|\eta|<2.5$.
Muons are matched to the primary vertex by requiring the longitudinal impact
parameter $z_0$ to satisfy $|z_0 \sin(\theta)|<0.5$ mm, where $\theta$ is the polar angle of the track.
The significance of the transverse impact parameter $d_0$ calculated with respect to the measured 
beam line position is required to satisfy  $|d_0|/\sigma(d_0) < 3$, where $\sigma(d_0)$ is the uncertainty in $d_0$.
Furthermore, the ``loose'' isolation criteria described in Ref.~\cite{Aad:2016jkr} are applied to suppress muons from $b$-hadron decays.
Jets are reconstructed using the anti-$k_t$ algorithm~\cite{antikt_algorithm} with a radius  parameter of $R=0.4$.
Candidate jets must have $|\eta|<4.5$, and the jet \pT must be larger than 25~(30)~\GeV\ for $|\eta|<$ 2.5 (2.5 $<|\eta|<$ 4.5).
To suppress pileup contributions, an additional requirement using the track and vertex information inside a jet~\cite{Aad:2015ina}
is imposed on jets with $|\eta|<2.4$ and $\pT < 60$~\GeV.
Top quark production is the second largest background with neutrinos and $b$ hadrons in the final states. 
Jets containing $b$ hadrons with $|\eta|<$ 2.5 are identified as $b$-tagged jets using a multivariate $b$-tagging algorithm 
that provides a 60\% efficiency and a rejection factor of more than 1000 for light-flavor jets~\cite{Aad:2015ydr}.
Neutrinos escape from the detector and lead to missing transverse momentum  \etmiss. 
The \etmiss is defined as the magnitude of the negative vectorial sum of 
the transverse momenta of the selected and calibrated physics objects (including muons and jets)
and the ID tracks not associated with any physics object (soft term)~\cite{Aad:2016nrq}.
To reduce the top quark contribution, events are required to have \etmiss $<$~80~\GeV\ and no $b$-tagged jets.

\begin{figure}[!htbp]
\begin{center}
\includegraphics[width=0.49\textwidth]{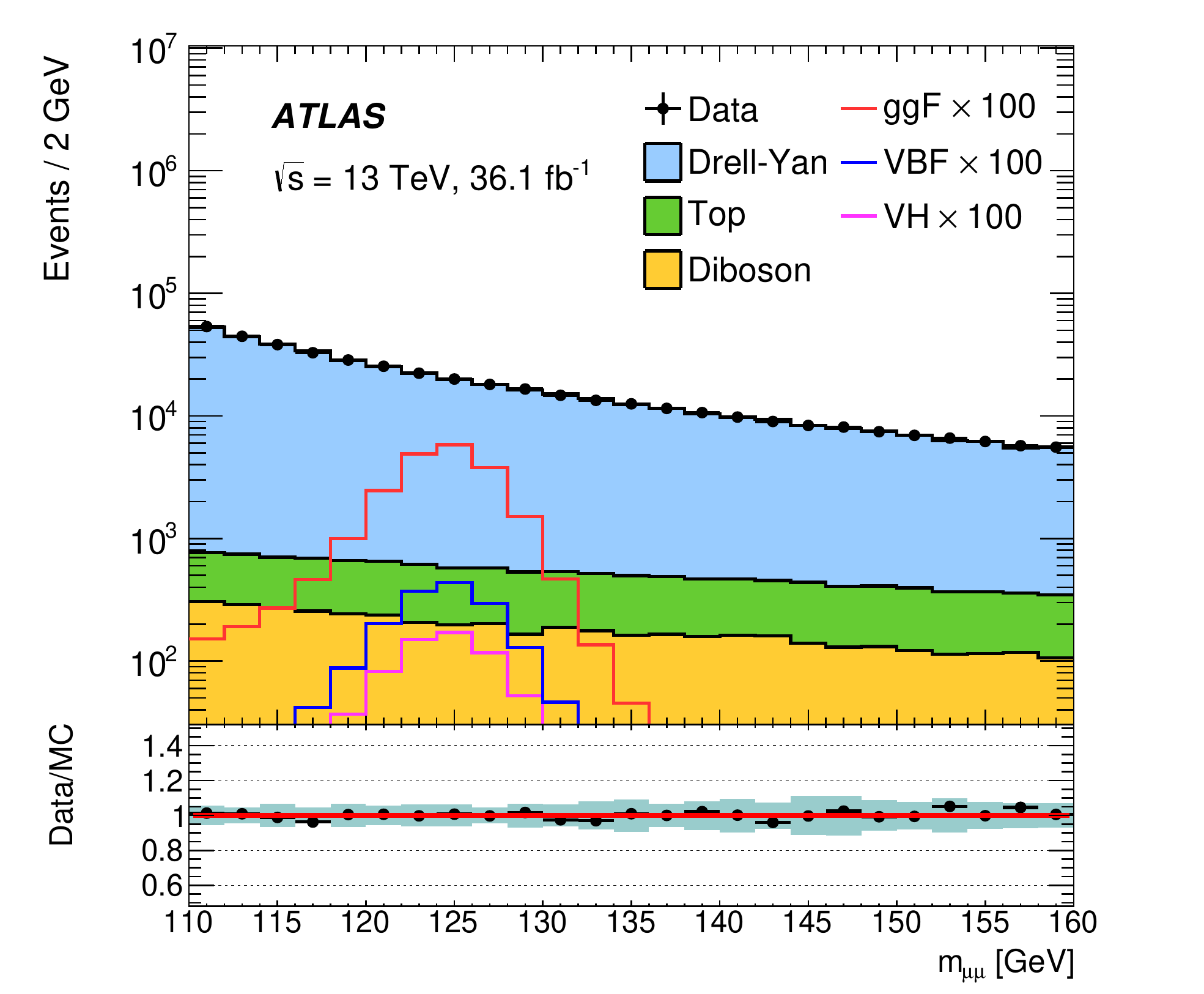}
\end{center}
  \caption{
  Observed and simulated $m_{\mu\mu}$ distributions in the inclusive signal region.
  The expected signals are scaled by a factor of 100.
  The total background prediction is normalized to the observed data yield, 
  while the relative fractions between the different processes are fixed to the SM predictions.
  The error band only reflects the statistical and experimental uncertainties 
  in the MC background prediction, while the theoretical uncertainties are not included.
}
\label{fig:evt:mumu}
\end{figure}

To ensure a high trigger efficiency, the leading muon must have $p_\mathrm{T}$ > 27~\GeV.
These criteria form the preselection, 
and events passing the preselection with $110~\GeV\ < m_{\mu\mu} < 160$~\GeV\ constitute the inclusive signal region. 
The signal efficiency is 57\% (59\%) for the \textit{gg}F (VBF) process. 
The $m_{\mu\mu}$ distributions for data and MC events in the inclusive signal region are shown in Figure~\ref{fig:evt:mumu}.

%Therefore, 
The VBF categories are only considered for events containing at least two jets.
To optimize the selections, several kinematic variables that are sensitive to 
the characteristics of the VBF production are used.
For jet-related variables, only the two jets with highest \pT are considered,
with the leading (subleading) jet denoted by $j_1$ ($j_2$).
Among those variables, the most sensitive ones are dijet invariant mass ($m_{jj}$),  \ptmm, difference in 
pseudorapidity $\Delta\eta_{jj}$, and angular distance $\Delta R_{jj}$ between the two jets.
Other variables with less discriminating power include transverse momentum of the dijet system ($p_{\mathrm{T}}^{jj}$), 
\etmiss, scalar \pT sum of muons and jets ($S_{\mathrm{T}}$), \pT of the system containing two muons 
and one or two jets ($p_{\mathrm{T}}^{\mu\mu j_1}$, $p_{\mathrm{T}}^{\mu\mu j_2}$ and 
$p_{\mathrm{T}}^{\mu\mu jj}$), rapidity difference between the dimuon system and the jets
($\Delta y_{\mu\mu, j_1}$, $\Delta y_{\mu\mu, j_2}$ and $\Delta y_{\mu\mu, jj}$), and 
``centrality'', 
defined as the difference between the dimuon rapidity and the averaged jet rapidity divided by the absolute rapidity difference between $j_1$ and $j_2$.
The MC modeling of these variables for the Drell-Yan process is compared with data in the region with 76~\GeV\ < $m_{\mu\mu}$ < 106~\GeV,
and no significant mismodeling is found.
All these variables are combined into a multivariate discriminant, 
which is then trained using MC events with a boosted-decision-tree (BDT) method~\cite{BDT,Roe:2004na,Hocker:2007ht} 
to maximize the separation between the VBF signal and the total background.
Events with a larger BDT score are more signallike, while background events tend to populate the low BDT score region.
Finally, events with BDT score $\geq$ 0.9 constitute one of the VBF categories (``VBF tight''), 
and the other one (``VBF loose'') is defined with $0.7<\text{BDT score}<0.9$.

The remaining events that are not selected for the VBF categories all enter into the \textit{gg}F categories.
Signal events from the \textit{gg}F process tend to have a harder \ptmm spectrum than Drell-Yan events
due to the higher initial-state QCD radiation.
To take advantage of this feature, events are separated into three \ptmm categories: ``low \ptmm'' (\ptmm $\leq 15$~\GeV),
``medium \ptmm'' (15~\GeV\ $<$ \ptmm $<$~50~\GeV), and ``high \ptmm'' (\ptmm $\geq$~50~\GeV).
Since the  muon momentum resolution in the barrel region ($|\eta| \leq 1.05$) is better than that in the end cap regions ($1.05<|\eta|<2.7$), 
events in each \pT category are further divided according to the pseudorapidities of the muons.
Requiring both muons to have $|\eta| \leq 1$ forms the ``central'' category, while the remaining events constitute
the ``noncentral'' category. 

Table~\ref{tab:evt:evtyield}
shows the expected signal and background event yields as well as the observed number of data events
within an $m_{\mu\mu}$ interval in each category. 
Each chosen interval is centered at the simulated signal peak and contains 90\% of the expected signal events.
These numbers are provided to demonstrate the expected detection sensitivity,
while in the final results, the signal and background yields are determined 
by fitting the observed $m_{\mu\mu}$ distributions.
\begin{table}[!htbp]
\begin{center}
\scalebox{0.9}{
    \begin{tabular}{|l|r|r|r|c|r|r|r|}
    \hline
                               &        $S$ &        $B$ &  $S$/$\sqrt{B}$ &            FWHM (\GeV) &       Data \\    \hline
             Central low \ptmm &         11 &       8000 &            0.12 &        5.6  &       7885 \\ %\hline
          Noncentral low \ptmm &         32 &      38000 &            0.16 &        7.0  &      38777 \\ %\hline
          Central medium \ptmm &         23 &       6400 &            0.29 &        5.7  &       6585 \\ %\hline
       Noncentral medium \ptmm &         66 &      31000 &            0.37 &        7.1  &      31291 \\ %\hline
            Central high \ptmm &         16 &       3300 &            0.28 &        6.3  &       3160 \\ %\hline
         Noncentral high \ptmm &         40 &      13000 &            0.35 &        7.7  &      12829 \\ %\hline
                     VBF loose &        3.4 &        260 &            0.21 &        7.6  &        274 \\ %\hline
                     VBF tight &        3.4 &         78 &            0.38 &        7.5  &         79 \\ \hline
        \end{tabular}
}
\end{center}
        \caption{
        Event yields for the expected signal ($S$) and background ($B$) processes,
        and numbers of the observed data events in different categories.
        The full widths at half maximum (FWHM) of the signal $m_{\mu\mu}$ distributions are also shown.
        In each category, the event yields are counted within an $m_{\mu\mu}$ interval,
        which is centered at the simulated signal peak and contains 90\% of the expected signal events.
        The expected signal event yields are normalized to \ulumi.
        The background in each category is normalized to the observed data yield, 
        while the relative fractions between the different processes are fixed to the SM predictions.
         }
        \label{tab:evt:evtyield}
\end{table}

Analytical models are used to describe the $m_{\mu\mu}$ distributions for both the signal and background processes. 
To describe the  Higgs boson peak with a lower-mass tail due to final-state photon radiation, 
the signal model is chosen as the sum of a Crystal Ball function (CB)~\cite{Oreglia:1980cs} and a Gaussian function (GS):
\begin{equation*}
\begin{split}
  P_{\rm S}(m_{\mu\mu})~=~& f_{\rm CB} \times {\rm CB}(m_{\mu\mu}, m_{\rm CB}, \sigma_{\rm CB}, \alpha, n) \\ 
& + (1-f_{\rm CB})\times {\rm GS} \left(m_{\mu\mu}, m_{\rm GS}, \sigma_{\rm GS}^{\rm S}\right),
\end{split}
\end{equation*} 
where $f_{\rm CB}$ is the fraction of the CB contribution when each component (CB or GS) is normalized to unity. 
The parameters $\alpha$ and $n$ define the power-law tail of the CB distribution. 
The parameters $m_{\rm CB}$, $m_{\rm GS}$, $\sigma_{\rm CB}$, and $\sigma_{\rm GS}^{\rm S}$ 
denote the CB mean value, GS mean value,  CB width, and GS width, respectively.
These parameters are determined for each signal category by fitting
the signal model to the simulated $m_{\mu\mu}$ spectrum. 
In each category, the \textit{gg}F, VBF, and $VH$ signal shapes are obtained separately 
and then combined into the total signal shape according to their SM predictions.

The background model should be able to describe the steeply falling $m_{\mu\mu}$ distributions from the dominant Drell-Yan process.
At the same time, it should have sufficient flexibility to absorb potential differences between data and MC simulation, 
and allow variations in the $m_{\mu\mu}$ spectra due to different selections and 
additional contributions from minor background processes.
The adopted model is the sum of a Breit-Wigner function (BW) convolved with a GS, and an exponential function divided by
a cubic function:
\begin{equation*}
 \begin{split}
P_{\rm B}(m_{\mu\mu})~=~& f \times [{\rm BW}(m_{\rm BW}, \Gamma_{\rm BW}) \otimes  {\rm GS}(\sigma_{\rm GS}^{\rm B})](m_{\mu\mu}) \\
& + (1-f) \times {\rm e}^{A\cdot m_{\mu\mu}}/m_{\mu\mu}^{3},
\end{split}
\end{equation*} 
where $f$ is the fraction of the BW component when each component is normalized to unity. 
The $\sigma_{\rm GS}^{\rm B}$ parameter in each category is fixed to 
the corresponding average $m_{\mu\mu}$ resolution as determined from MC Drell-Yan events. 
For all the categories, the BW parameters are fixed to $m_{\rm BW}=91.2$~\GeV\ and $\Gamma_{\rm BW} = 2.49$~\GeV~\cite{Olive:2016xmw}. 
The parameters $f$ and $A$ are unconstrained and uncorrelated between different categories.

A binned maximum-likelihood fit to the observed $m_{\mu\mu}$ distributions in the range \frange 
is performed using the sum of the signal and background models (``S+B model''). 
The fit is done simultaneously in all the categories. 
In addition to the background model parameters ($f$ and A) described earlier, 
the background normalization in each category is a free parameter in the fit. 
The product of the \hmm signal strength $\mu_{\rm S}$ and the expected signal yield gives the signal normalization in each category.

The expected signal yields used in the fit
are subject to experimental and theoretical uncertainties.
The systematic uncertainties in the expected signal are correlated between all the categories.

The uncertainty in the combined 2015 and 2016 integrated luminosity is 3.2\%,
derived, following a methodology similar to that detailed in Ref.~\cite{Aaboud:2016hhf}, 
from a preliminary calibration of the luminosity scale using $x$--$y$ beam-separation scans performed in August 2015 and May 2016.
Other sources of  experimental uncertainty include the muon reconstruction and identification efficiencies, 
the efficiencies due to the trigger, isolation and impact parameter requirements, 
the muon momentum scale and resolution, the determination of the \etmiss soft term, 
the $b$-tagging efficiency, 
the pileup modeling, as well as the jet energy scale and resolution.
The total experimental uncertainty in the predicted signal yield in each \textit{gg}F category 
is between 4\% and 6\%, dominated by the luminosity, muon, jet, and pileup contributions.
The experimental uncertainty increases to 15\% in the VBF categories, 
due to larger contributions from the jet energy scale and resolution uncertainties.
The effects of the experimental uncertainties in the predicted signal $m_{\mu\mu}$ shapes are found to be minor and are therefore neglected in this search.

The theoretical uncertainties in the production cross section of the Higgs boson and 
the \hmm decay branching ratio are set according to Refs.~\cite{Dittmaier:2011ti,Dittmaier:2012vm}.
The uncertainty in the signal acceptance in the \textit{gg}F categories, due to the modeling of the Higgs boson \pt\ spectrum,
is estimated by varying the QCD scales used in the \textsc{HRes} program.
The acceptance uncertainties of \textit{gg}F signal events in the VBF categories are estimated using the method described in Ref.~\cite{Aad:2014xva}. 
The uncertainties associated with the modeling of multiparton interactions are estimated by turning them off in the event generation,
according to the recommendations in Ref.~\cite{Heinemeyer:2013tqa}.
The uncertainty in the \textit{gg}F signal prediction ranges from 15\% to 25\%, 
dominated by the uncertainties due to omitted high-order effects.
The total theoretical uncertainty in the VBF signal yield in each category is typically around 5\%.

Any systematic bias in the background model when describing the underlying $m_{\mu\mu}$ spectrum
might result in spurious signal events in the measurement. 
In each category, the number of spurious signal events ($N_{\rm spur}$) is 
estimated by fitting the parameterized S+B model 
to the simulated background $m_{\mu\mu}$ distribution in the range 110--160~\GeV.
The $m_{\mu\mu}$ spectra are obtained from large Drell-Yan MC samples,
which were produced with \textsc{Powheg-box} v2~\cite{Frixione:2007vw} and \textsc{MadGraph}5~\cite{Alwall:2011uj}  
for the \textit{gg}F and VBF categories, respectively, 
and correspond to an equivalent integrated luminosity of about 5 ab$^{-1}$.
Values of $N_{\rm spur}$ are derived for three nearby Higgs boson masses (120, 125, and 130~\GeV),
and from these the largest value between the yields and their statistical uncertainties is taken as the $N_\mathrm{spur}$ value for a certain category.
A detailed discussion about how $N_\mathrm{spur}$ is used in the fitting procedure is given in Ref.~\cite{Aad:2014eha}.
The background modeling uncertainty is treated as uncorrelated among all the categories. 
This uncertainty varies from 8\% to 50\% of the statistical uncertainties of the background, 
depending on the selection category.
The impact of the background mismodeling on the expected upper limit 
on the signal strength is about 2\%.

\begin{figure}[!htbp]
 \begin{center}
  \includegraphics[width=0.45\textwidth]{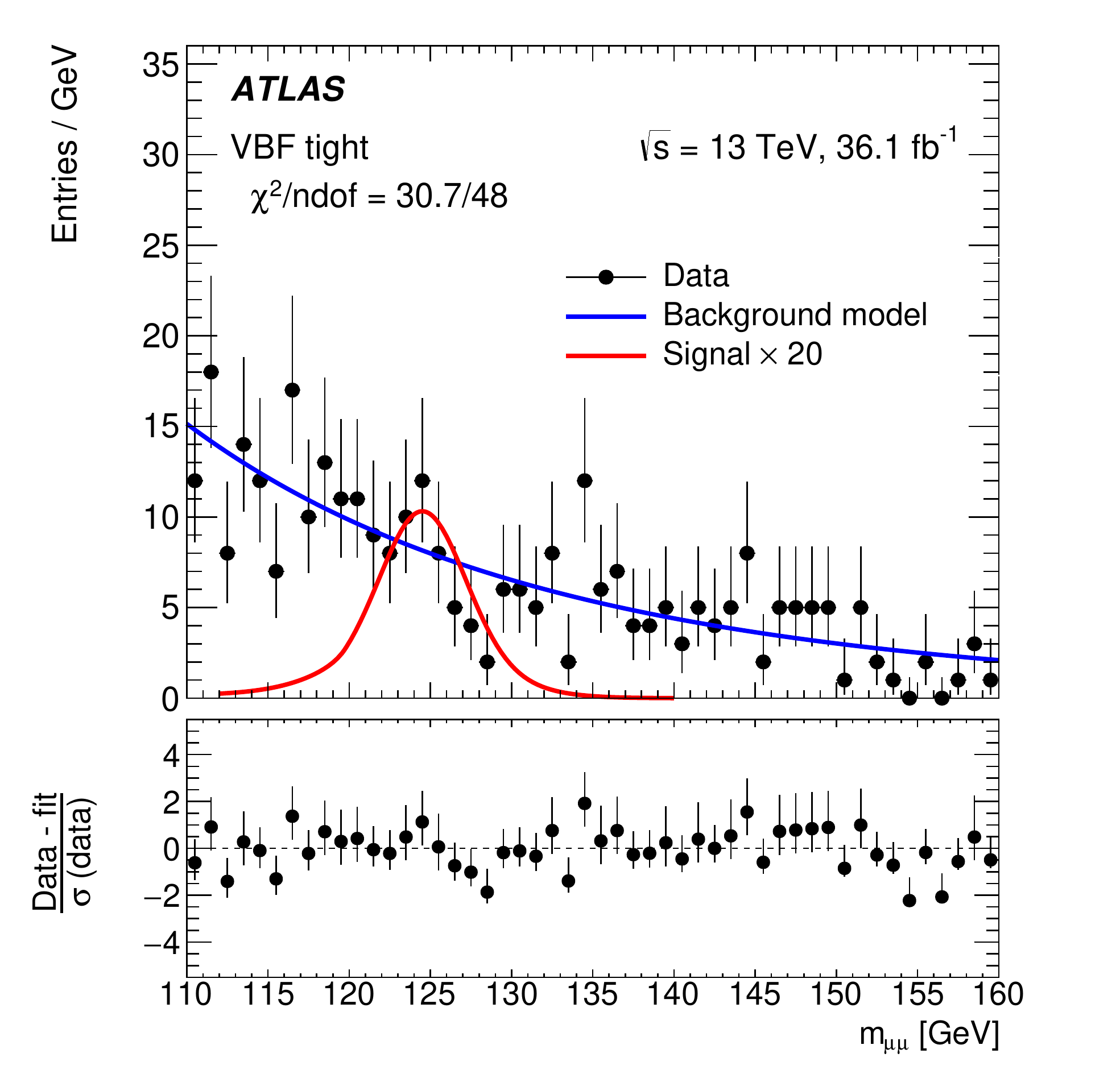}
\end{center}
  \caption{Background-only fit to the observed $m_{\mu\mu}$ distribution 
    in the VBF tight category. 
    Only the statistical uncertainties are shown for the data points.
    The expected signal is scaled by a factor of 20.
   \label{fig:fit:VBFtight}
  }
\end{figure}

The observed $m_{\mu\mu}$ spectrum is compared to the background-only fit in Figure~\ref{fig:fit:VBFtight} for the VBF tight category.
The S+B model is fitted to the observed $m_{\mu\mu}$ spectra in eight signal categories simultaneously,
and the measured overall signal strength is $\mu_{\rm S}  = -0.1\pm 1.5$. 
An upper limit on $\mu_{\rm S}$ is computed using a modified frequentist CL$_\text{s}$ method~\cite{Cowan:2010st,Read:2002hq} with the 
profile-likelihood-ratio test statistic~\cite{Cowan:2010st}.
The observed (expected) upper limit on $\mu_{\rm S}$ at the 95\% C.L. is found to be 3.0~(3.1). 
 This limit is driven by the data statistical uncertainty, 
 while the impact of the systematic uncertainties is found to be 2.2\%.
When combined with the ATLAS Run 1 data, 
the observed (expected) upper limit is 2.8~(2.9) at the 95\% C.L..
The corresponding measured signal strength is $\mu_{\rm S}  = -0.1\pm 1.4$.
The theoretical and experimental uncertainties in the expected signal  and the background modeling uncertainty
are correlated in the combination.

To conclude, a search for the dimuon decay of the Higgs boson is performed using 
\ulumi of data collected with the ATLAS detector in $pp$ collisions at $\sqrt{s}=$13~\TeV\ at the LHC.
No significant excess is observed in data, and an upper limit is set on the signal strength.
%The uncertainty on the signal strength has decreased by 44\% over the previous combined ATLAS and CMS measurement~\cite{Khachatryan:2016vau}. 

% Acknowledgements
% Version 6-Mar-2017
%----------------------------------------------
We thank CERN for the very successful operation of the LHC, as well as the
support staff from our institutions without whom ATLAS could not be
operated efficiently.

We acknowledge the support of ANPCyT, Argentina; YerPhI, Armenia; ARC, Australia; BMWFW and FWF, Austria; ANAS, Azerbaijan; SSTC, Belarus; CNPq and FAPESP, Brazil; NSERC, NRC and CFI, Canada; CERN; CONICYT, Chile; CAS, MOST and NSFC, China; COLCIENCIAS, Colombia; MSMT CR, MPO CR and VSC CR, Czech Republic; DNRF and DNSRC, Denmark; IN2P3-CNRS, CEA-DSM/IRFU, France; SRNSF, Georgia; BMBF, HGF, and MPG, Germany; GSRT, Greece; RGC, Hong Kong SAR, China; ISF, I-CORE and Benoziyo Center, Israel; INFN, Italy; MEXT and JSPS, Japan; CNRST, Morocco; NWO, Netherlands; RCN, Norway; MNiSW and NCN, Poland; FCT, Portugal; MNE/IFA, Romania; MES of Russia and NRC KI, Russian Federation; JINR; MESTD, Serbia; MSSR, Slovakia; ARRS and MIZ\v{S}, Slovenia; DST/NRF, South Africa; MINECO, Spain; SRC and Wallenberg Foundation, Sweden; SERI, SNSF and Cantons of Bern and Geneva, Switzerland; MOST, Taiwan; TAEK, Turkey; STFC, United Kingdom; DOE and NSF, United States of America. In addition, individual groups and members have received support from BCKDF, the Canada Council, CANARIE, CRC, Compute Canada, FQRNT, and the Ontario Innovation Trust, Canada; EPLANET, ERC, ERDF, FP7, Horizon 2020 and Marie Sk{\l}odowska-Curie Actions, European Union; Investissements d'Avenir Labex and Idex, ANR, R{\'e}gion Auvergne and Fondation Partager le Savoir, France; DFG and AvH Foundation, Germany; Herakleitos, Thales and Aristeia programmes co-financed by EU-ESF and the Greek NSRF; BSF, GIF and Minerva, Israel; BRF, Norway; CERCA Programme Generalitat de Catalunya, Generalitat Valenciana, Spain; the Royal Society and Leverhulme Trust, United Kingdom.

The crucial computing support from all WLCG partners is acknowledged gratefully, in particular from CERN, the ATLAS Tier-1 facilities at TRIUMF (Canada), NDGF (Denmark, Norway, Sweden), CC-IN2P3 (France), KIT/GridKA (Germany), INFN-CNAF (Italy), NL-T1 (Netherlands), PIC (Spain), ASGC (Taiwan), RAL (UK) and BNL (USA), the Tier-2 facilities worldwide and large non-WLCG resource providers. Major contributors of computing resources are listed in Ref.~\cite{ATL-GEN-PUB-2016-002}.
%----------------------------------------------

\bibliographystyle{atlasBibStyleWithTitle}
\bibliography{HmumuPaper}

\begin{thebibliography}{10}
\bibitem{AtlasCoordFootnote}
ATLAS uses a right-handed coordinate system with its origin at the nominal
  interaction point (IP) in the center of the detector and the $z$ axis along
  the beam pipe. The $x$ axis points from the IP to the center of the LHC ring,
  and the $y$ axis points upwards. Cylindrical coordinates $(r,\phi)$ are used
  in the transverse plane, $\phi$ being the azimuthal angle around the $z$
  axis. The pseudorapidity is defined in terms of the polar angle $\theta$ as
  $\eta = -\ln \tan(\theta/2)$. Angular distance is measured in units of
  $\Delta R \equiv \sqrt{(\Delta\eta)^{2} + (\Delta\phi)^{2}}$.
\end{thebibliography}

\newpage 
\ifprl 
   \onecolumngrid 
\fi
% ATLAS Collaboration author list
% Data extracted on 15-Apr-2017 for paper reference HIGG-2016-10
%\documentclass[11pt]{article}
%\usepackage{a4wide}
%\begin{document}
\begin{flushleft}
{\Large The ATLAS Collaboration}

\bigskip

M.~Aaboud$^\textrm{\scriptsize 137d}$,
G.~Aad$^\textrm{\scriptsize 88}$,
B.~Abbott$^\textrm{\scriptsize 115}$,
O.~Abdinov$^\textrm{\scriptsize 12}$$^{,*}$,
B.~Abeloos$^\textrm{\scriptsize 119}$,
S.H.~Abidi$^\textrm{\scriptsize 161}$,
O.S.~AbouZeid$^\textrm{\scriptsize 139}$,
N.L.~Abraham$^\textrm{\scriptsize 151}$,
H.~Abramowicz$^\textrm{\scriptsize 155}$,
H.~Abreu$^\textrm{\scriptsize 154}$
% ...
\bigskip
\\
% ...
$^{12}$ Institute of Physics, Azerbaijan Academy of Sciences, Baku, Azerbaijan\\
% ...
$^{88}$ CPPM, Aix-Marseille Universit{\'e} and CNRS/IN2P3, Marseille, France\\
% ...
$^{115}$ Homer L. Dodge Department of Physics and Astronomy, University of Oklahoma, Norman OK, United States of America\\
% ...
$^{119}$ LAL, Univ. Paris-Sud, CNRS/IN2P3, Universit{\'e} Paris-Saclay, Orsay, France\\
% ...
$^{137}$ $^{(a)}$ Facult{\'e} des Sciences Ain Chock, R{\'e}seau Universitaire de Physique des Hautes Energies - Universit{\'e} Hassan II, Casablanca; $^{(b)}$ Centre National de l'Energie des Sciences Techniques Nucleaires, Rabat; $^{(c)}$ Facult{\'e} des Sciences Semlalia, Universit{\'e} Cadi Ayyad, LPHEA-Marrakech; $^{(d)}$ Facult{\'e} des Sciences, Universit{\'e} Mohamed Premier and LPTPM, Oujda; $^{(e)}$ Facult{\'e} des sciences, Universit{\'e} Mohammed V, Rabat, Morocco\\
% ...
$^{139}$ Santa Cruz Institute for Particle Physics, University of California Santa Cruz, Santa Cruz CA, United States of America\\
% ...
$^{151}$ Department of Physics and Astronomy, University of Sussex, Brighton, United Kingdom\\
% ...
$^{154}$ Department of Physics, Technion: Israel Institute of Technology, Haifa, Israel\\
$^{155}$ Raymond and Beverly Sackler School of Physics and Astronomy, Tel Aviv University, Tel Aviv, Israel\\
% ...
$^{161}$ Department of Physics, University of Toronto, Toronto ON, Canada\\
% ...
$^{a}$ Also at Department of Physics, King's College London, London, United Kingdom\\
$^{b}$ Also at Institute of Physics, Azerbaijan Academy of Sciences, Baku, Azerbaijan\\
$^{c}$ Also at Novosibirsk State University, Novosibirsk, Russia\\
$^{d}$ Also at TRIUMF, Vancouver BC, Canada\\
$^{e}$ Also at Department of Physics {\&} Astronomy, University of Louisville, Louisville, KY, United States of America\\
% ...
$^{*}$ Deceased
\end{flushleft}

%\end{document}
% Created with xml2latex.py

\end{document}